\documentclass[%
 reprint,
 amsmath,amssymb,
 aps,
]{revtex4-2}

\usepackage{graphicx}
\usepackage{dcolumn}
\usepackage{bm}

\graphicspath{{images/}}
\newcommand{\customfigure}[3]{\begin{figure}%
\centering
\includegraphics[scale=#1]{#2.pdf}%
\centering%
\caption{#3}\label{fig:#2}%
\end{figure}}

\newcommand{\customfigurestar}[3]{\begin{figure*}%
\centering
\includegraphics[scale=#1]{#2.pdf}%
\centering%
\caption{#3}\label{fig:#2}%
\end{figure*}}

\begin{document}

\preprint{APS/123-QED}

\title{Resolving the evolution of natural fragment shapes}

\author{Balázs Havasi-Tóth}
 \email{havasi-toth.balazs@gpk.bme.hu}
 \affiliation{Department of Fluid Mechanics, Budapest University of Technology \\
 HUN-REN-BME Morphodynamics Research Group, Budapest}

\author{Eszter Fehér}
\affiliation{
Department of Morphology and Geometric Modeling, Budapest University of Technology \\
HUN-REN-BME Morphodynamics Research Group, Budapest
}%


\date{\today}

\begin{abstract}

We propose a geometrically motivated mathematical model which reveals the key features of coastal and fluvial fragment shape evolution from the earliest stages of the abrasion. Our \textit{collisional polygon model} governs the evolution through an ordinary differential equation (ODE) that determines the rounding rate of initially sharp corners in the function of the size reduction. As an approximation, the basic structure of our model adopts the concept of Bloore's partial differential equation (PDE) in terms of the curvature dependent local collisional frequency. We tested our model under various conditions and made comparisons with the predictions of Bloore's PDE. Moreover, we applied the model to discover and quantify the mathematical conditions corresponding to typical and special shape evolution. By further extending our model to investigate the self-dual and mixed cases, we outline a possible explanation of the long-term preservation of initial pebble shape characteristics.
\end{abstract}


\maketitle


\section{\label{sec:level1}Introduction}

The shape of rocks and sediment particles provides insight into their history and origin. Fragmentation processes create fragments that travel in fluvial and coastal environments from one place to another while hitting each other and the bedrock. The energy of these encounters, the size of the participating objects, and the travel distance all contribute to the geometry of the particle. We know from field and laboratory measurements, that large energy collision results in the sharp corners breaking off and the fragments turning into rounded pebbles. 

Mathematical theories exist that model the abrasion process as the evolution of the particle contour. Aristotle proposed a distance-driven model \cite{krynine1960antiquity}, in which the points on the contour move in the in-ward normal direction proportionally to their distance from the centroid. Building on the work of Rayleigh \cite{rayleigh1942ultimate, strutt1944pebbles}, Firey \cite{Firey1974} introduced a PDE to model the collision of particles with large surfaces by defining the speed of the points on the contour in the inward normal directions as:  
 \begin{equation} \label{eq:Firey}
    v=c\kappa,
\end{equation}
where $\kappa$ is the curvature and $c$ is a constant. It was proven by Gage \cite{Gage1983,Gage1986} and Grayson \cite{Grayson1987}, that the limit shape of (\ref{eq:Firey}) is a circle. Bloore's model \cite{Bloore1977} abandons the condition on the size of the abraders and allows for arbitrary-sized particles. In two dimensions, it reads as:
\begin{equation} \label{eq:Bloore}
    v=1+c\kappa.
\end{equation}
The constant $c$ depends on the relative size of the colliding particles \cite{Domokos2009,Varkonyi2011}. In the limit of $c=0$, (\ref{eq:Bloore}) reduces to the Eikonal equation describing the collision with infinitely small particles (e.g. sandblasting).  

Although Bloore's model is believed to accurately predict the shape evolution of almost arbitrary initial shapes, its vastness makes it very difficult to analyze its general behavior. Despite the sphere being an attractor of the three-dimensional collision models \cite{Firey1974,Andrews1999,Andrews2013,Bloore1977,Gage1983}, natural pebbles seem to never reach a spherical shape. In an attempt to explain this observation, many alternatives of Bloore's model and other mathematical theories were formulated considering additional effects such as friction and the motion patterns of the particles \cite{Szabo2015,Domokos2012,Hill2022,Winzer2013,Winzer2021,Winzer2023}.

In order for the underlying dynamics to be better understood, simplifications and subsystems of the equations were already investigated. Domokos and Gibbons \cite{Domokos2012} introduced a system of ODEs, namely the box equations that describe the evolution of highly abraded particles. We know from \cite{Domokos2014}, that the abrasion of fragments takes place in two phases: a rounding and a shrinking phase; as a result, the box equations approximate the evolution of the second phase.

Here, we introduce an ODE that describes the shape evolution of fragments from the very beginning of the abrasion process. We consider n-fold symmetric convex polygonal fragment shapes that collide with circular and convex n-fold symmetric rounded polygonal objects. We show that different conditions allow the identification of homothetic and other distinctive solutions by analytical expressions. It is also capable of modeling the self-dual case $-$ where the identical abraders equally participate in the collective abrasion process $-$ and a mixed one, where the self-dual behavior is diluted with very small abraders. Remarkably, the non-monotonic rounding of the mixed case led to the discovery of the existence of non-trivial homothetic evolution.
The paper is organized as follows: in Section \ref{sec:model}, we derive the model for circular and fragment abraders, and verify it by the numerical simulation of (\ref{eq:Bloore}). We show applications of the model in Section \ref{sec:case_studies}.

\section{The collisional polygon model}\label{sec:model}
\subsection{Circular abraders}\label{sec:circular}
Following Bloore's thought process, the collisional frequency with isotropically distributed circular abraders attacking the perimeter of an arbitrary convex shape is defined as
\begin{equation} \label{eq:collisional_frequency}
    f_c(t)=1+R^*\kappa(t)=\frac{R(t)+R^*}{R(t)},
\end{equation}
where $R(t)=1/\kappa(t)$ is the radius of the local osculating circle of the closed perimeter described with the parameter $t$, and $R^*$ is the radius of the circular abrader. Apparently, the frequency rapidly increases when $R(t)\rightarrow 0$, and becomes unity along flat sides. However, the frequency is hard to be expressed in a generic case, which provides the ultimate reason of the high complexity of (\ref{eq:Bloore}). Therefore, we propose a simplification on $R(t)$ in the present section by formulating suitable restrictions on the geometric evolution. 

We consider the two-dimensional shape evolution of regular n-fold symmetric polygons ($n\geq3$) with rounded corners representing the geometries of natural fragment shapes after various degrees of abrasion. The three parameters chosen here for the clear definition of such a shape are the inscribed circle's diameter $a$, the interior angle $\varphi=\pi(n-2)/n$ between any two adjacent sides, and the arc radius $R\leq a/2$ at the corners. Any fragment shape determined by $a$, $\varphi$ and $R$ reduces (\ref{eq:collisional_frequency}) to a piecewise constant function as
\begin{equation} \label{eq:probability}
f_c=
\begin{cases}
  1 & \text{along the straight edges,}\\
  \frac{R+R^*}{R} & \text{along the arcs.}\\
\end{cases}
\end{equation}

In Figure \ref{fig:drawing}, a single corner of an arbitrary fragment is presented with the corresponding notations. The shrinking of the perimeter is qualitatively illustrated with the red arrows. We approximate the fragment shapes with rounded polygons during the abrasion; accordingly, we hypothetically assume that the circular arcs remain circular and the straight edges remain straight. After an infinitesimally small step $\mathrm{d}a$, the shape is defined by $a-\mathrm{d}a$ and $R+\mathrm{d}R$, while $\varphi=const$. Note that $\mathrm{d}a$ and $\mathrm{d}R$ correspond to the size reduction and rounding of the fragment, respectively. Utilizing that the abrasion speed at the straight and curved segments is proportional with the collisional frequency, we compute the abrasion speed $\nu$ at the midpoints of the arcs as
\begin{equation}\label{eq:rho}
    \nu=\frac{R+R^*}{R}\frac{\mathrm{d}a}{2},
\end{equation}
where $\mathrm{d}a/2$ is the abrasion speed of the straight edges.

\customfigure{0.25}{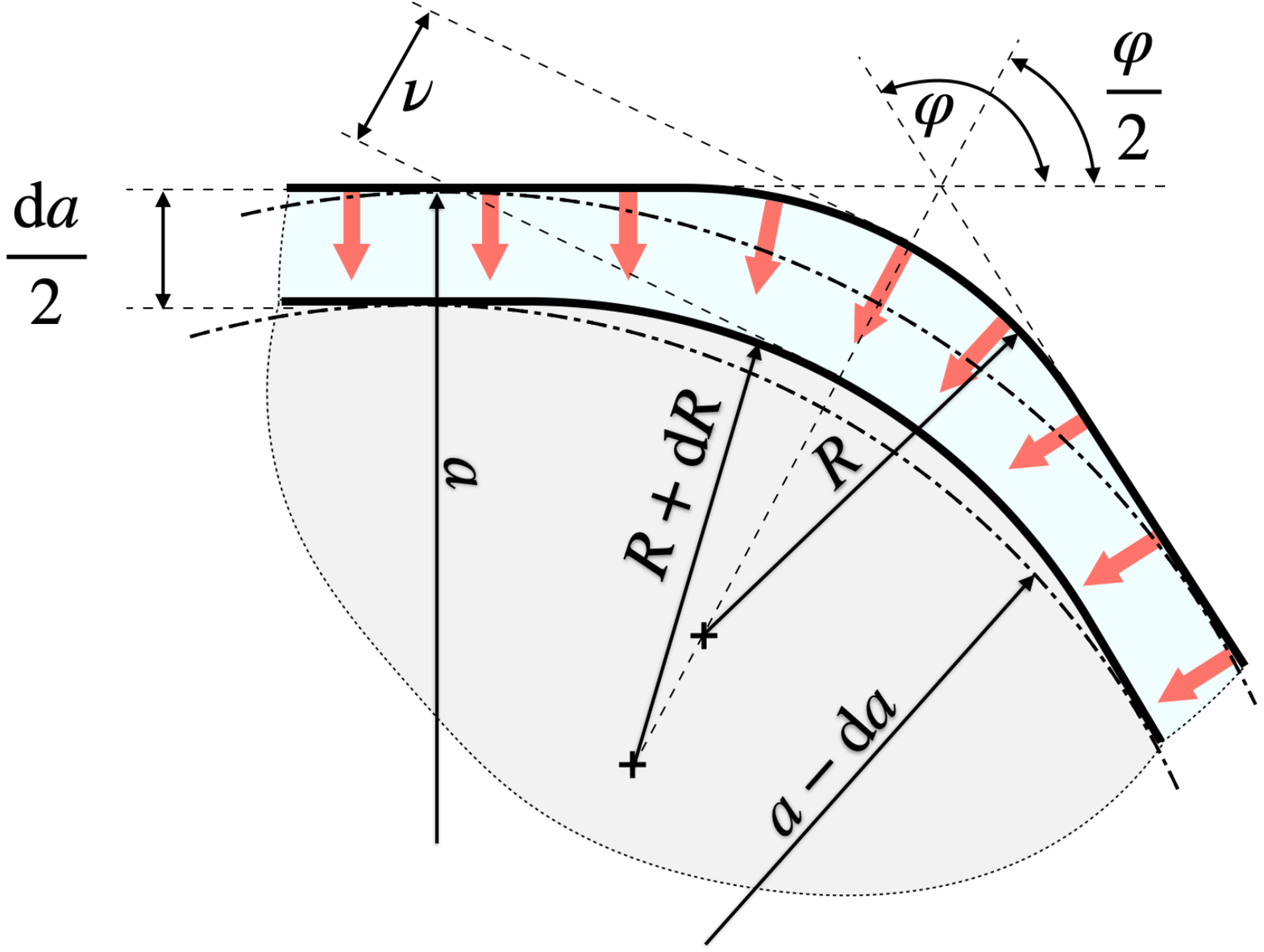}{Schematic explanation of the infinitesimal step between two consecutive shapes of a rounded polygon characterized by $a$, $R$ and $\varphi$. The movement of the contour in the inward direction is illustrated by the red arrows. Light bluish areas correspond to the worn material, gray color indicates the new shape. }

From geometrical considerations based on Figure \ref{fig:drawing}, the value of $\nu$ is also expressed geometrically so that
\begin{widetext}
\begin{equation} \label{eq:geom_and_speed}
    -\nu=\frac{R+\mathrm{d}R}{\sin\big(\frac{\varphi}{2}\big)}-(R+\mathrm{d}R)-\frac{\mathrm{d}a}{2\sin\big(\frac{\varphi}{2}\big)}-\frac{R}{\sin\big(\frac{\varphi}{2}\big)}+R=\mathrm{d}R\bigg(\frac{1}{\sin\big(\frac{\varphi}{2}\big)}-1\bigg)-\frac{\mathrm{d}a}{2\sin\big(\frac{\varphi}{2}\big)},
\end{equation}
\end{widetext}
where the last term on the right hand side implies the displacement of the center of the arc. Combining (\ref{eq:rho}) and (\ref{eq:geom_and_speed}), a first order ordinary differential equation is obtained as
\begin{equation} \label{eq:ode1}
    \frac{\mathrm{d}R}{\mathrm{d}a}=\frac{R\bigg(1-\frac{1}{\sin\big(\frac{\varphi}{2}\big)}\bigg)+R^*}{2R\bigg(1-\frac{1}{\sin\big(\frac{\varphi}{2}\big)}\bigg)},
\end{equation}
which determines the kinematics of the radius $R$ as $a$ decreases. The effects of the controlling parameter $R^*$ will be investigated later in Sections \ref{sec:fragment_abraders} and \ref{sec:case_studies}, but from this point, we consider a temporally constant abrader radius if not indicated otherwise, i.e., the original shape of the abrader is preserved during the process.

The solution of (\ref{eq:ode1}) is constructed analytically, with the initial condition
\begin{equation}
    R(a_0)=R_0,
\end{equation}
where $a_0$ is the inscribed circle's diameter of the initial fragment shape. The solution in implicit form then yields
\begin{widetext}
\begin{equation} \label{eq:solution} 
    a=a_0+2R-2R_0+\frac{2SR^*\Big[\ln\big(-R_0+S(R_0+R^*)\big)-\ln\big(-R+S(R+R^*)\big)\Big]}{S-1},
\end{equation}
\end{widetext}
where $S=\sin\big(\varphi/2\big)$ is substituted for better readability. Without the intention of a parameter reduction of (\ref{eq:ode1}), we can choose $a_0=1$ for the sake of simplicity. In the current section, initially sharp fragments are taken into account with the initial condition $R_0=0$.

\customfigure{0.305}{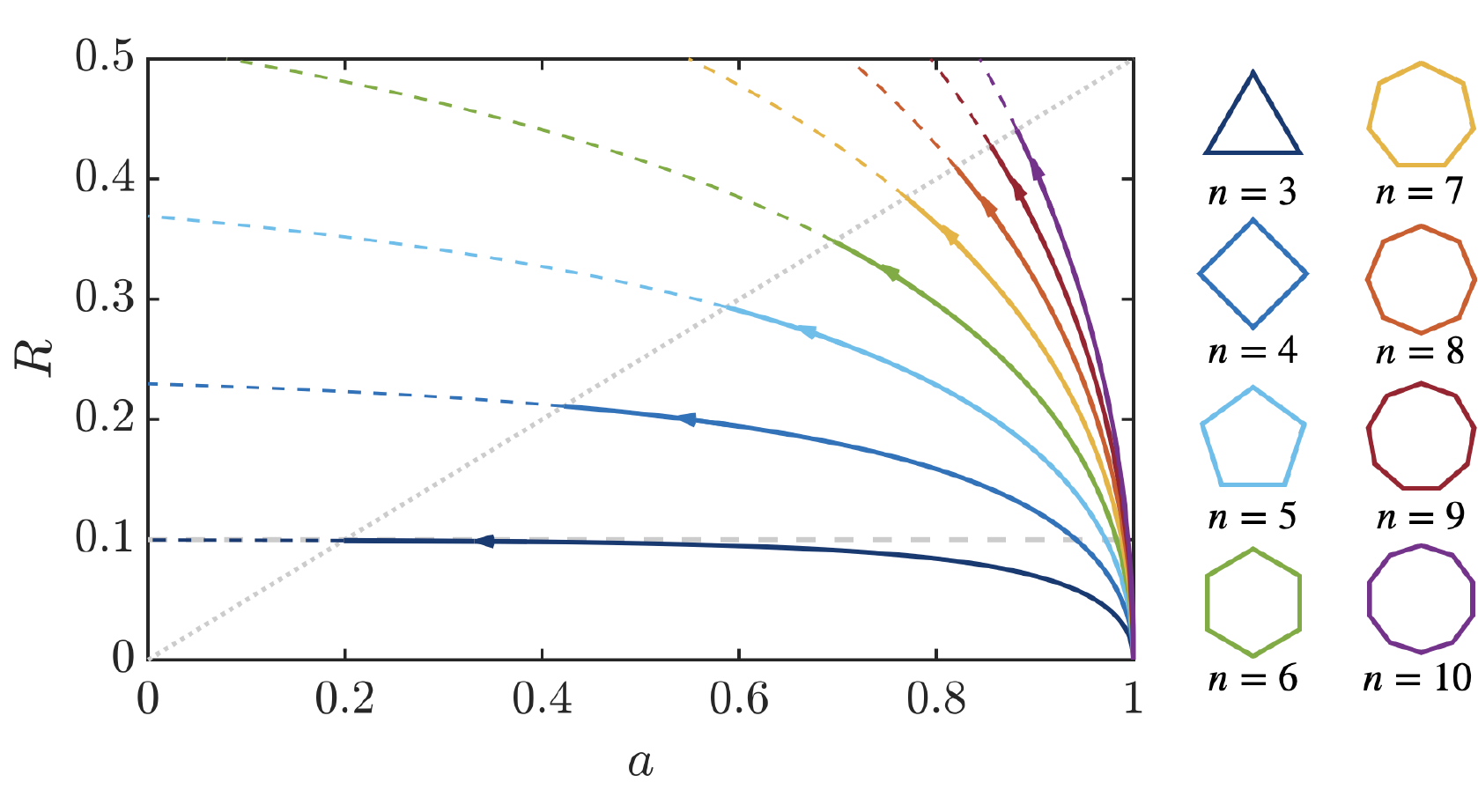}{Solutions of (\ref{eq:ode1}) in the case of polygonal fragments with sharp corners and $n=3..10$, while $R^*=0.1$. Horizontal and diagonal dashed lines mark the $R=R^*$ and $R=a/2$ lines, respectively and arrows indicate the direction of the evolution. Solutions above the $R=a/2$ line correspond to geometrically infeasible polygons and marked with dashed lines.}

\customfigurestar{0.72}{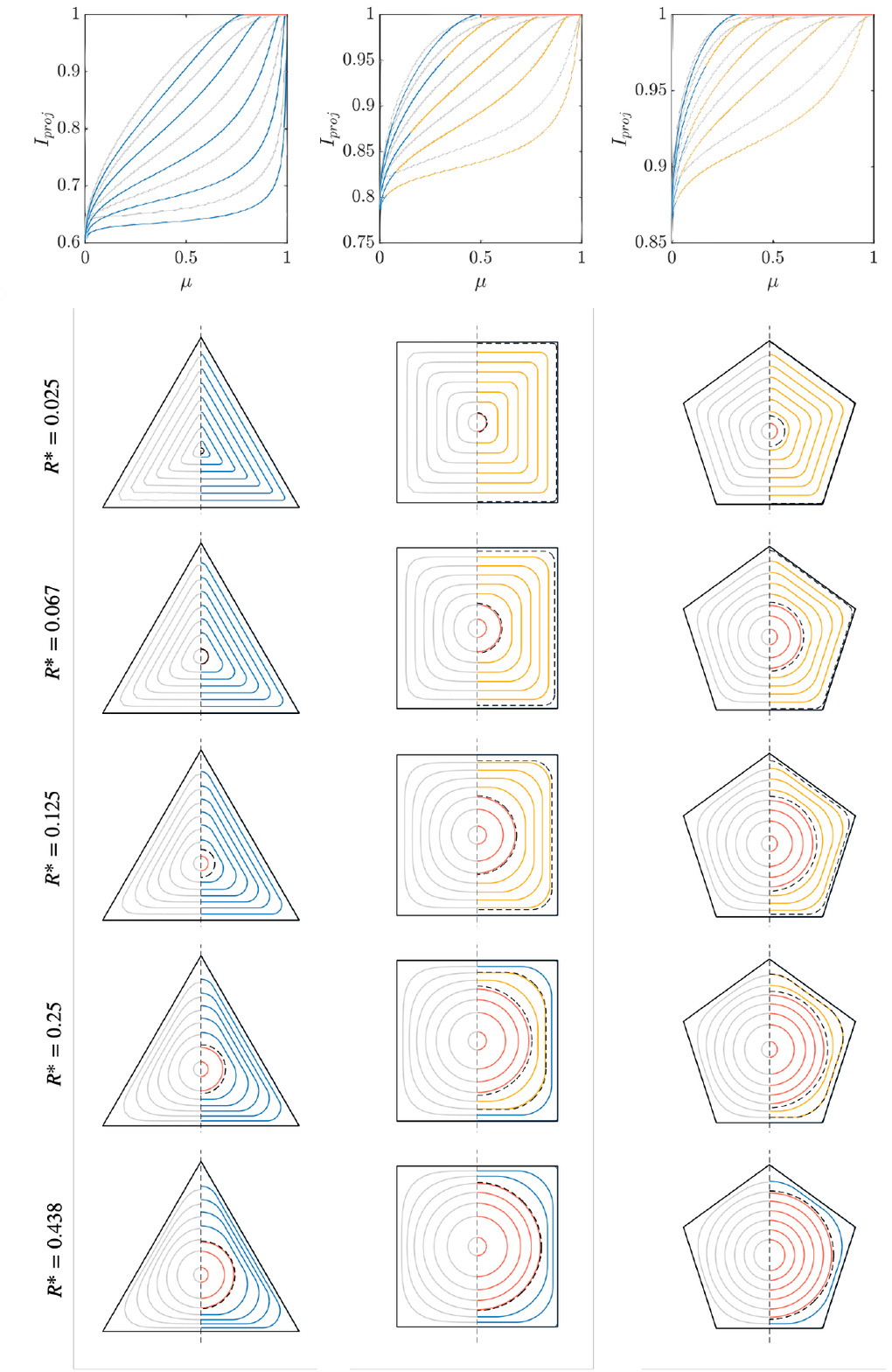}{Comparison of the collisional polygon model with the numerical solution of the PDE (\ref{eq:Bloore}) in case of $n=3,4,5$ and different abrader radii $R^*$. Gray contours (left halves) and $I_{proj}$ curves mark the solutions of (\ref{eq:Bloore}), the color-coded ones were obtained using the collisional polygon model solution (\ref{eq:solution}).}

By plotting (\ref{eq:solution}), Figure \ref{fig:Ra_n3456} shows how the radius increases as the polygon shrinks under collisional abrasion with circular abraders. As soon as the rounding fragments pass through the circular shape at $R=a/2$, the corresponding polygons become geometrically infeasible and marked with the immediate transition from solid to dashed lines. It should also be pointed out that after reaching the circular shape, each trajectory should follow the attractor $R=a/2$ until both $R$ and $a$ become zero. Since the model is valid until the shape reaches the circle, we only investigate the kinematics below $R=a/2$, and we assume that the evolution after reaching $R=a/2$ consists of a sequence of shrinking circles through a trivially homothetic evolution. Less straightforward homothetic cases are investigated later in Section \ref{sec:homothetic} and \ref{sec:self_dual_mixed}.

In agreement with our expectations, higher $n$ values result in faster convergence (cf. higher slope) to the circle $R=a/2$. Apart from the triangle ($n=3$), all polygons' $R$ values reach the abrader's radius $R^*$ at a certain point before reaching the circular shape. Furthermore, in agreement with \citet{Firey1974,Bloore1977}, the model suggests circular limit shapes when $R^*>0$
\begin{equation}
    \frac{\mathrm{d}R}{\mathrm{d}a}=\frac{R\bigg(1-\frac{1}{\sin\big(\frac{\varphi}{2}\big)}\bigg)+R^*}{2R\bigg(1-\frac{1}{\sin\big(\frac{\varphi}{2}\big)}\bigg)}<\frac{1}{2}, \\
\end{equation}
and the vertical slope at $R(a_0)=0$ (sharp corners) as
\begin{equation}
    \lim_{R\to 0} \frac{\mathrm{d}R}{\mathrm{d}a}=\lim_{R\to 0} \frac{R\bigg(1-\frac{1}{\sin\big(\frac{\varphi}{2}\big)}\bigg)+R^*}{2R\bigg(1-\frac{1}{\sin\big(\frac{\varphi}{2}\big)}\bigg)}=-\infty. \\
\end{equation}

Using the solution (\ref{eq:solution}), the evolution of initially sharp fragments with $n=3,4$ and $5$ is shown in Figure \ref{fig:all_evolution} with different circular abrader radii. Besides the predictions of the model, we plotted the light gray contours of the PDE ($\ref{eq:Bloore}$) with the same initial shapes and $R^*$ values. In the top row, the two-dimensional
\begin{equation}
    I_{proj.}=\frac{4\pi A}{P^2},
\end{equation}
isoperimetric ratios are compared in all cases for both models, where $A$ and $P$ are the area and perimeter of the fragment respectively. Colors of the contours and the corresponding $I_{proj.}$ values are chosen during the evolution as follows:
\begin{itemize}
    \item blue: $R<R^*$,
    \item yellow: $R^*<R<a/2$,
    \item red: $R=a/2$,
\end{itemize}
which regions are separated with the dashed lines between the contours.
As expected, the yellow region was never reached when $n=3$, and the same stands for the square and pentagon shapes when the
\begin{equation}
    R^*>R(a)=\frac{a}{2}
\end{equation}
condition holds.

Despite being and ODE, the contours show very good agreement in terms of the shape evolution provided by the PDE. Although our model consistently underestimates the isoperimetric ratios, a qualitative match can be claimed with a relative error of $<10\%$ with respect to the difference between the initial and final values of $I_{proj}$.

Since the properties of the abrader particles are described solely by $R^*$ in (\ref{eq:ode1}), the rest of the paper focuses on different, geometrically and geologically motivated choices of $R^*$ values and functions.

\subsection{Fragment abraders} \label{sec:fragment_abraders}
Throughout the previous subsection, the abrasion in a constant environment of circular abraders was considered. To take polygonal abraders into account, similarly to \citet{Domokos2009}, we propose an average radius $\bar R^*$ computed for any regular convex polygon with arbitrary arc radius determined by the values $a^*$, $R^*$ and $\varphi^*$.

We can formulate the average radius as
\begin{equation} \label{eq:polygon_abrader_proportion}
    \frac{R+\bar R^*}{R}=1+\frac{\kappa}{\bar\kappa^*},
\end{equation}
where $\kappa$ is the curvature of the arcs at the corners of the abraded polygon, and the bar represents spatial averaging over the perimeter of the abrader. Considering the perimeter of a polygonal abrader with rounded corners as
\begin{equation}
    P^*=2L^*n+R^*2\pi=2n^*\cot\bigg(\frac{\varphi^*}{2}\bigg)\bigg(\frac{a^*}{2}-R^*\bigg)+R^*2\pi,
\end{equation}
the average curvature yields
\begin{equation}
\begin{split}
    \bar \kappa^*&=\frac{2n^*L^*\cdot 0+R^*2\pi\frac{1}{R^*}}{P^*}=\\
    &=\frac{2\pi}{2n^*\cot\big(\frac{\varphi^*}{2}\big)\big(\frac{a^*}{2}-R^*\big)+R^*2\pi},
\end{split}
\end{equation}
where $L^*$ is the length of the straight edges. Thus the average radius of the abrader becomes
\begin{equation} \label{eq:average_polygon_radius}
    \bar R^*=\frac{1}{\bar\kappa^*}=R^*+\frac{2\cot\big(\frac{\varphi^*}{2}\big)}{(\pi-\varphi^*)}\bigg(\frac{a^*}{2}-R^*\bigg),
\end{equation}
which is a single constant value for any given fragment shape with arbitrary corner radius.
Substituting (\ref{eq:average_polygon_radius}) with the constant $R^*$ in the original differential equation (\ref{eq:ode1}) we obtain
\begin{equation} \label{eq:ode2}
    \frac{\mathrm{d}R}{\mathrm{d}a}=\frac{R\bigg(1-\frac{1}{\sin\big(\frac{\varphi}{2}\big)}\bigg)+\bar R^*}{2R\bigg(1-\frac{1}{\sin\big(\frac{\varphi}{2}\big)}\bigg)},
\end{equation}
for which the construction of the analytical solution can be done the same way as before, since $\bar R^*$ is constant. Moreover, (\ref{eq:ode2}) evidently reduces to (\ref{eq:ode1}) if $a^*=2R^*$, being the condition of a circular abrader. Considering the abundance of (\ref{eq:ode2}) in terms of possible variations of the abrader and abraded sizes and shapes, we only derived this model as a prerequisite for the self-dual case in Section \ref{sec:self_dual_mixed} section and omit its further investigations in the present work.

\section{Application case studies and results} \label{sec:case_studies}

\customfigurestar{0.55}{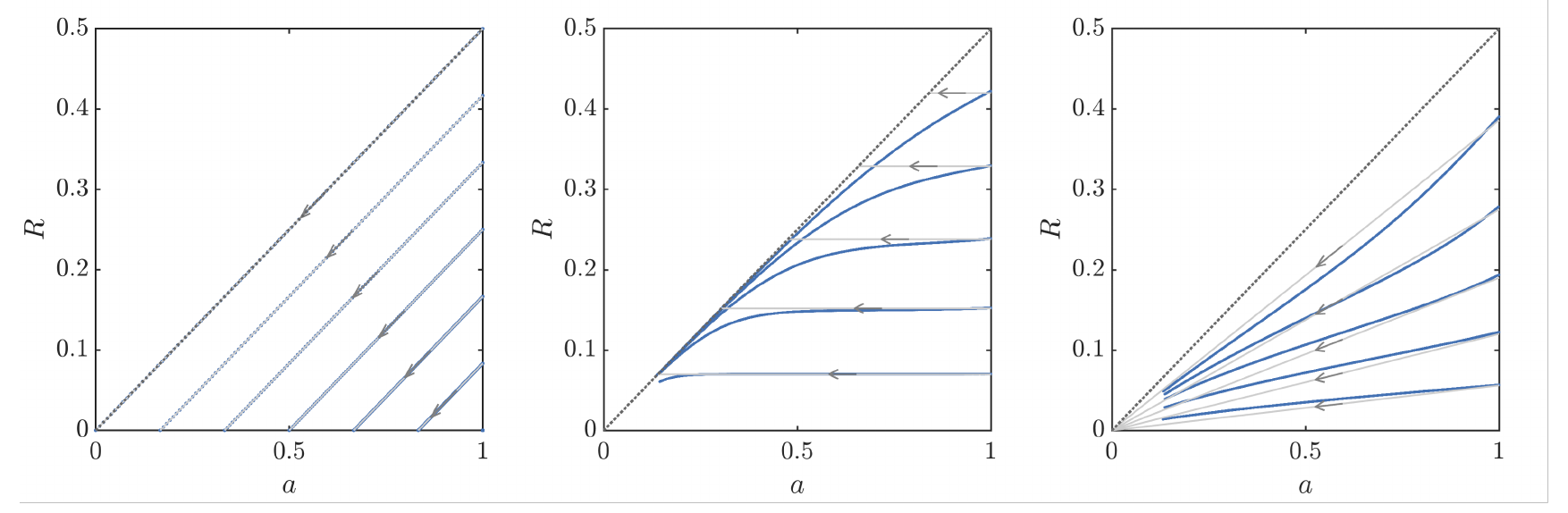}{Flows obtained by the presented special constraints. Left to right: abrasion by dust, stationary radius $R$, and the homothetic evolution. Blue and gray lines show the solutions of the full PDE and the flow predicted by our model.}

This section lists a few relevant findings of our collisional polygon model in terms of the special cases of the evolution; we focus on the characteristics of the fragment shape evolution and make attempts to predict the required abrader properties that control the processes.

In order for the evolution presented in this section to be comparable with the numerical solution of Bloore's equation, we define the $a$ and $R$ values for any closed planar curve $\rho(\theta)$ as
\begin{equation} \label{eq:bloore_Ra}
\begin{split}
    &a=2\rho_{min} \\
    &R=\frac{1}{\kappa_{max}},
\end{split}
\end{equation}
where the maximum curvature $\kappa_{max}$ is the curvature at $\rho_{max}$. Using the definitions (\ref{eq:bloore_Ra}), the trajectories of the full PDE become representable in the $R-a$ plane.

\subsection{Heterogeneous environment}
We have shown in Section \ref{sec:fragment_abraders} that an equivalent circular abrader corresponds to any fragment abrader. The implication of multiple different radii $R^*_i$ of the abraders is considered here to present the applicability to heterogeneous environments. Following the idea of Domokos and V\'arkonyi \cite{Domokos2009}, we aim to show how a single $\hat R^*$ value can represent an arbitrary set of abraders in our model.

In the case of two different abraders, we define $p$ and $q=1-p$ to be the probability of the collision with an abrader of radius $R^*_1$ and $R^*_2$ respectively. Using (\ref{eq:geom_and_speed}), the proportion can be written as
\begin{widetext}
\begin{equation} \label{eq:multiple_abrader}
    \mathrm{d}R\bigg(\frac{1}{\sin\big(\frac{\varphi}{2}\big)}-1\bigg)+\frac{1}{\sin\big(\frac{\varphi}{2}\big)}=p\frac{R+R^*_1}{R}\frac{\mathrm{d}a}{2}+(1-p)\frac{R+R^*_2}{R}\frac{\mathrm{d}a}{2}=
    \frac{R+pR^*_1+(1-p)R^*_2}{R}\frac{\mathrm{d}a}{2}=\frac{R+\hat R^*}{R}\frac{\mathrm{d}a}{2},
\end{equation}
\end{widetext}
suggesting the single equivalent particle radius to be computed for an arbitrary collection as
\begin{equation}
    \hat R^*=\sum_i{p_iR^*_i},
\end{equation}
where $p_i$ is the probability of the collision with particle radius $R^*_i$ so that
\begin{equation}
    \sum_i{p_i}=1.
\end{equation}
    
\subsection{Abrasion by dust ($R^*=0$)} \label{sec:abrasion_by_dust}
The highest possible slope that produces decreasing radius $R$ can be obtained by setting the radius of the abraders to zero:
\begin{equation} \label{eq:abrasion_by_dust}
    \frac{\mathrm{d}R}{\mathrm{d}a}=\frac{R\bigg(1-\frac{1}{\sin\big(\frac{\varphi}{2}\big)}\bigg)+0}{2R\bigg(1-\frac{1}{\sin\big(\frac{\varphi}{2}\big)}\bigg)}=\frac{1}{2},
\end{equation}
which results in diagonal straight trajectories parallel with the $R=a/2$ line. Such evolution converges to sharp fragment shapes ($R=0$) in case of any initial condition $0<R_0<a_0/2$. It is evident, that after reaching the sharp polygonal shape, the evolution ultimately becomes trivially homothetic. Another consequence of (\ref{eq:abrasion_by_dust}) is that no feasible condition ($R^*\geq0$) exists for circular shapes to become non-circular. A less straightforward homothetic shape transition is discussed in Section \ref{sec:homothetic}.

The corresponding flow in the $R-a$ plane is presented in Figure \ref{fig:special_lines}a together with the solution of Bloore's equation. 

\subsection{Stationary radius $R$}
If the constraint
\begin{equation}
    \frac{\mathrm{d}R}{\mathrm{d}a}=\frac{R\bigg(1-\frac{1}{\sin\big(\frac{\varphi}{2}\big)}\bigg)+R^*}{2R\bigg(1-\frac{1}{\sin\big(\frac{\varphi}{2}\big)}\bigg)}=0,
\end{equation}
holds, one can find a stationary value of $R$ in the function of $R^*$ for any $\varphi$ as
\begin{equation} \label{eq:staticR_condition}
    R=\frac{R^*}{\frac{1}{\sin\big(\frac{\varphi}{2}\big)}-1},
\end{equation}
which explains why the radius of the triangle's arcs never reaches $R^*$. Moreover, as Figure \ref{fig:Ra_n3456} already suggested, the corner radii of any initially sharp regular convex polygon is unable to become constant during the evolution when $R^*$ is constant. Nevertheless, for a given $R$, we can express a required $R^*$ from (\ref{eq:staticR_condition}).

The comparison of the stationary $R$ flow with Bloore's equation is shown in Figure \ref{fig:special_lines}b.

\subsection{Homothetic evolution with $a/2>R>0$} \label{sec:homothetic}
In order to obtain a sequence of self-similar solutions, the constraint
\begin{equation} \label{eq:homothetic_constraint}
    \frac{\mathrm{d}R}{\mathrm{d}a}=\frac{R}{a}
\end{equation}
must be fulfilled. Together with (\ref{eq:ode1}), the required non-constant $R^*(R,a)$ function is given as
\begin{equation}
    R^*(R,a)=\frac{\bigg(\frac{1}{\sin\big(\frac{\varphi}{2}\big)}-1\bigg)R(a-2R)}{a}.
\end{equation}

The non-trivial homothetic flow together with the evolution proposed by Bloore is shown in Figure \ref{fig:special_lines}c.

These three analytically generated flows show qualitative similarities with the full PDE providing a chance of better understanding the underlying dynamics and conditions in terms of the abrasion environment.

\subsection{Self-dual and mixed cases} \label{sec:self_dual_mixed}
After the deduction of (\ref{eq:ode2}), a rich mathematical formulation is constructed for the shape evolution of arbitrary symmetric fragments and abraders of n-fold symmetry.

If the abraders are the exact copies of the abraded fragment and equally participate in the morphodynamic process, the abrasion occurs in a continuously changing environment. In other words, the abrasion becomes a collective process called the self-dual evolution, where the abraders and the abraded shapes are identical at every instant and none of the particles play a distinguished role. This restriction significantly reduces the degrees of freedom of (\ref{eq:ode2}) by setting
\begin{equation} \label{eq:self_dual_constraints}
\begin{split}
    \varphi^*\rightarrow\varphi &\iff n^*\rightarrow n, \\
    &a^*\rightarrow a, \\
    &R^*\rightarrow R,
\end{split}
\end{equation}
which is technically corresponding to a varying $R^*$.
Unfortunately, after the substitution of (\ref{eq:self_dual_constraints}), the analytical solution of (\ref{eq:ode2}) becomes cumbersome, therefore we hereinafter resort to a numerical analysis.

The system now reads as
\begin{equation} \label{eq:ode3}
    \frac{\mathrm{d}R}{\mathrm{d}a}=\frac{\bigg(1-\frac{1}{\sin\big(\frac{\varphi}{2}\big)}\bigg)+1+\frac{2\cot\big(\frac{\varphi}{2}\big)}{(\pi-\varphi)}\big(\frac{a}{2R}-1\big)}{2\bigg(1-\frac{1}{\sin\big(\frac{\varphi}{2}\big)}\bigg)},
\end{equation}
and the initial condition $R(a_0)=0$ is considered. Note that (\ref{eq:ode3}) now completely lacks of a controlling parameter. We solved the system using the fourth order Runge-Kutta method and visualized the solutions in Figure \ref{fig:Ra_n3456_selfdual}. Two significant conclusions can be made based on the figure. The trajectories in the $a-R$ plane became much steeper at the beginning, while on the other hand, the solutions became seemingly non-monotonic in the case of $n=3$ and $4$. We extended the vertical axis upper limit to present this difference between the trajectories. This phenomena supported our initial conjecture that the self-dual abrasion might be accountable for a less intuitive evolution.

\customfigure{0.305}{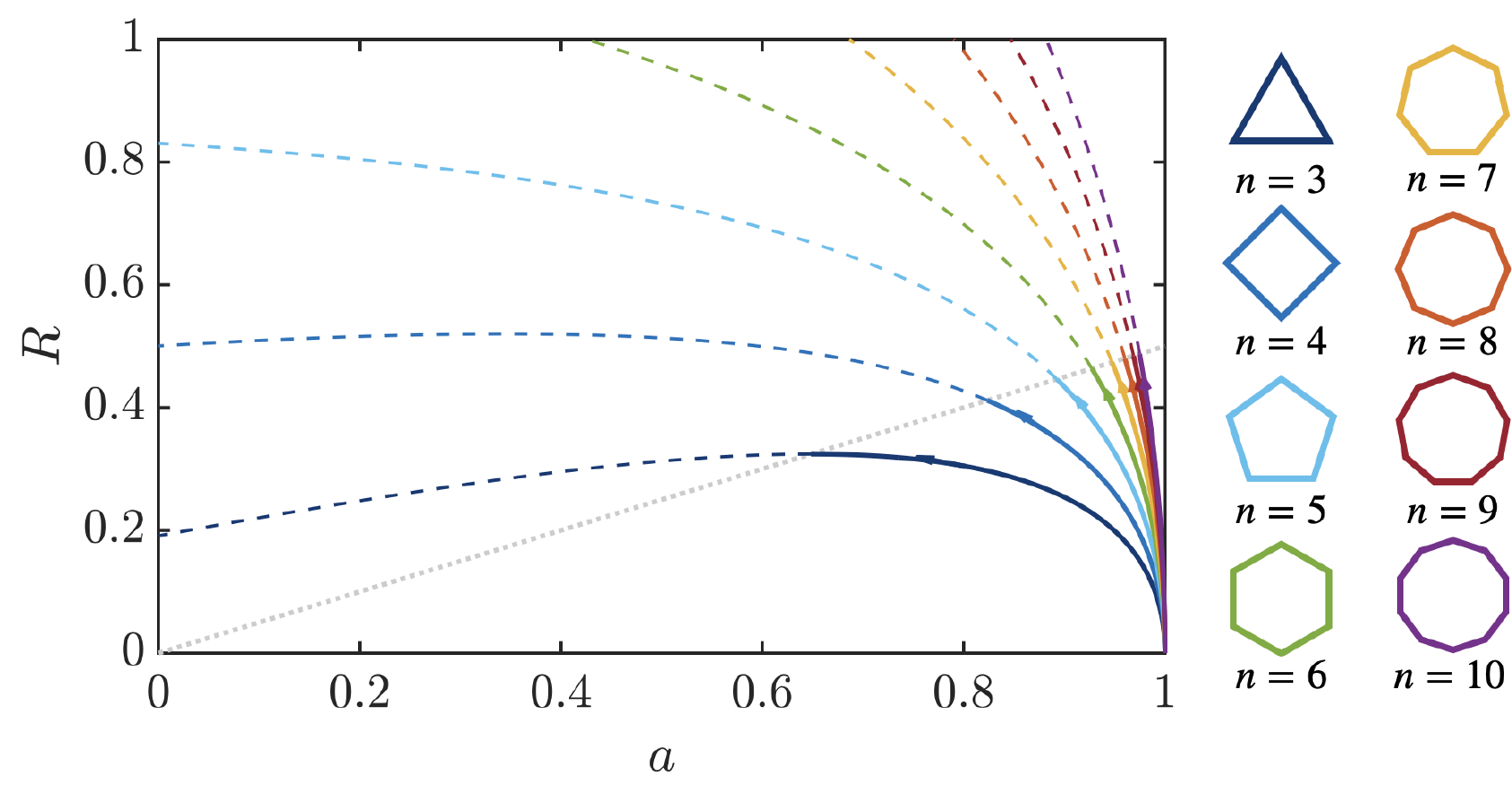}{Self-dual evolution of $n=3..10$ fragments using (\ref{eq:ode2}) with the constraints (\ref{eq:self_dual_constraints}). Diagonal line marks $R=a/2$ and arrows indicate the direction of the evolution. Solutions above the $R=a/2$ line correspond to geometrically infeasible polygons and marked with dashed lines. Vertical axis range was extended for better visibility.} 

We performed further computations with the same constraints but we also implied small abraders with $R^*=0$ with various collisional probability $1-p$. Here we use
\begin{equation}
    \hat R^*=p \bar R^*+(1-p)R^*=p\bar R^*,
\end{equation}
deduced from the formula of the heterogeneous environment (\ref{eq:multiple_abrader}). We can consider it as a dilution of the self-dual evolution, for which the complete system is expressed as
\begin{equation} \label{eq:ode4}
    \frac{\mathrm{d}R}{\mathrm{d}a}=\frac{\bigg(1-\frac{1}{\sin\big(\frac{\varphi}{2}\big)}\bigg)+p+p\frac{2\cot\big(\frac{\varphi}{2}\big)}{(\pi-\varphi)}\big(\frac{a}{2R}-1\big)}{2\bigg(1-\frac{1}{\sin\big(\frac{\varphi}{2}\big)}\bigg)}.
\end{equation}
\customfigure{0.305}{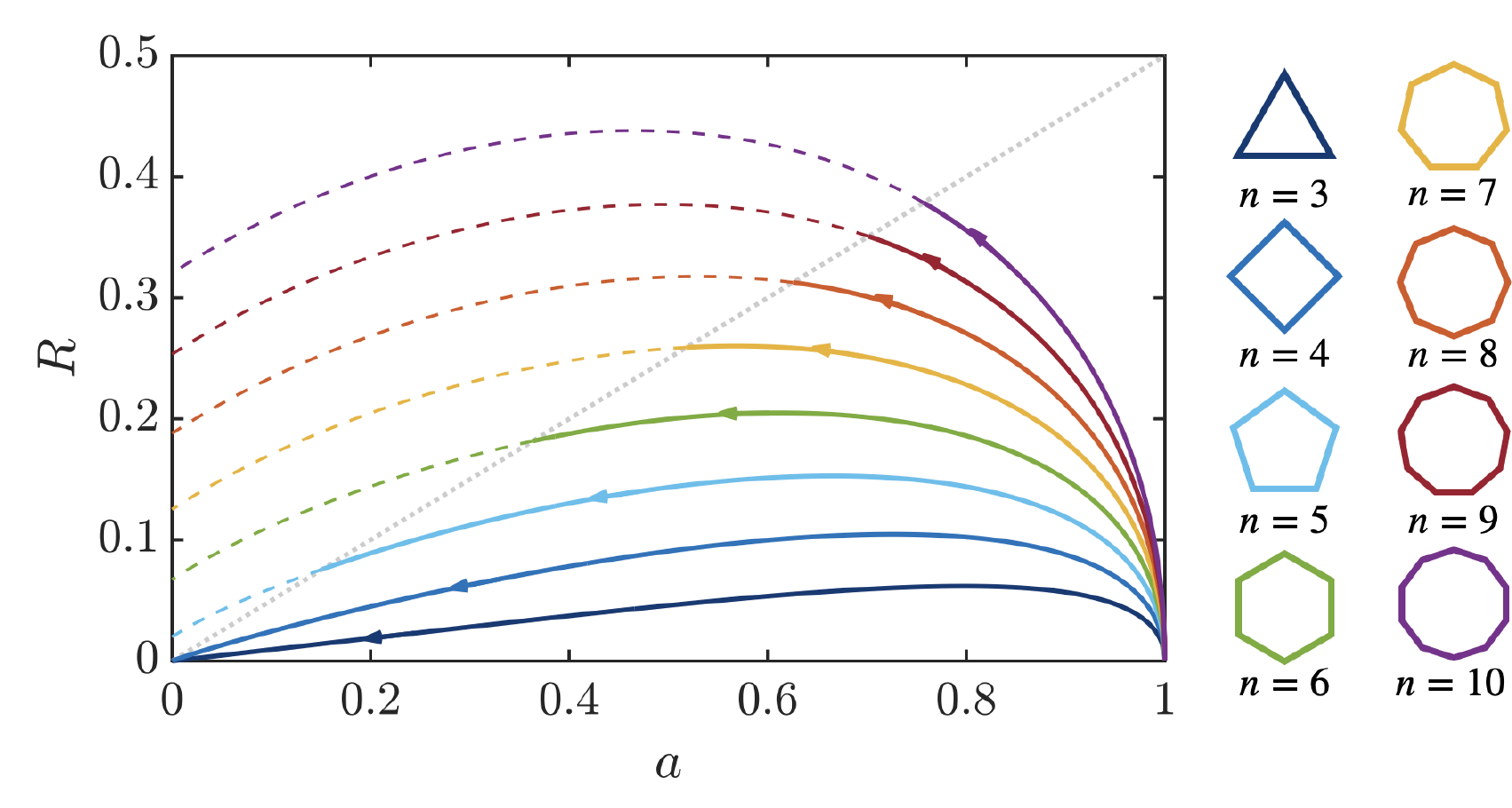}{Self-dual mixed evolution of $n=3..10$ fragments using (\ref{eq:ode2}) with the constraints (\ref{eq:self_dual_constraints}) and $p=0.1$. Diagonal line marks $R=a/2$ and arrows indicate the direction of the evolution. Solutions above the $R=a/2$ line correspond to geometrically infeasible polygons and marked with dashed lines.}
The trajectories of (\ref{eq:ode4}) with $p=0.1$ are presented in Figure \ref{fig:Ra_n3456_selfdual_mixed}, where the non-monotonic regions now appear in the area below $R=a/2$. Moreover, the lower curves corresponding to $n<5$ polygons seem to converge to straight lines in the vicinity of the origin, suggesting that homothetic evolution might exist for the mixed environment. Accordingly, we performed an analysis of the system (\ref{eq:ode3}) with the constraint of the homothetic evolution as
\begin{equation} \label{eq:ode4_homothetic}
    \frac{\mathrm{d}R}{\mathrm{d}a}=\frac{\bigg(1-\frac{1}{\sin\big(\frac{\varphi}{2}\big)}\bigg)+p+p\frac{2\cot\big(\frac{\varphi}{2}\big)}{(\pi-\varphi)}\big(\frac{1}{2\alpha}-1\big)}{2\bigg(1-\frac{1}{\sin\big(\frac{\varphi}{2}\big)}\bigg)}=\alpha,
\end{equation}
by looking for solutions of $\alpha=R/a$ in the function of $p$. The solid and dashed curves in Figure \ref{fig:selfdual_mixed_homothetic} show stable and unstable solutions of $\alpha$ in the function of $p$, respectively. In agreement with the intuition, higher order symmetry of the fragments requires higher rate of dilution (smaller $p$) of the self-dual evolution with the dust. In the case of no solution exists for a specific fragment and $p$ value, the evolution never becomes homothetic. However, our deduction suggests that a sufficiently high amount of dust might help to drive the self-dual evolution of fragments to a non-trivially homothetic one without directly forcing the corresponding constraint (\ref{eq:homothetic_constraint}) in the system.

\customfigure{0.305}{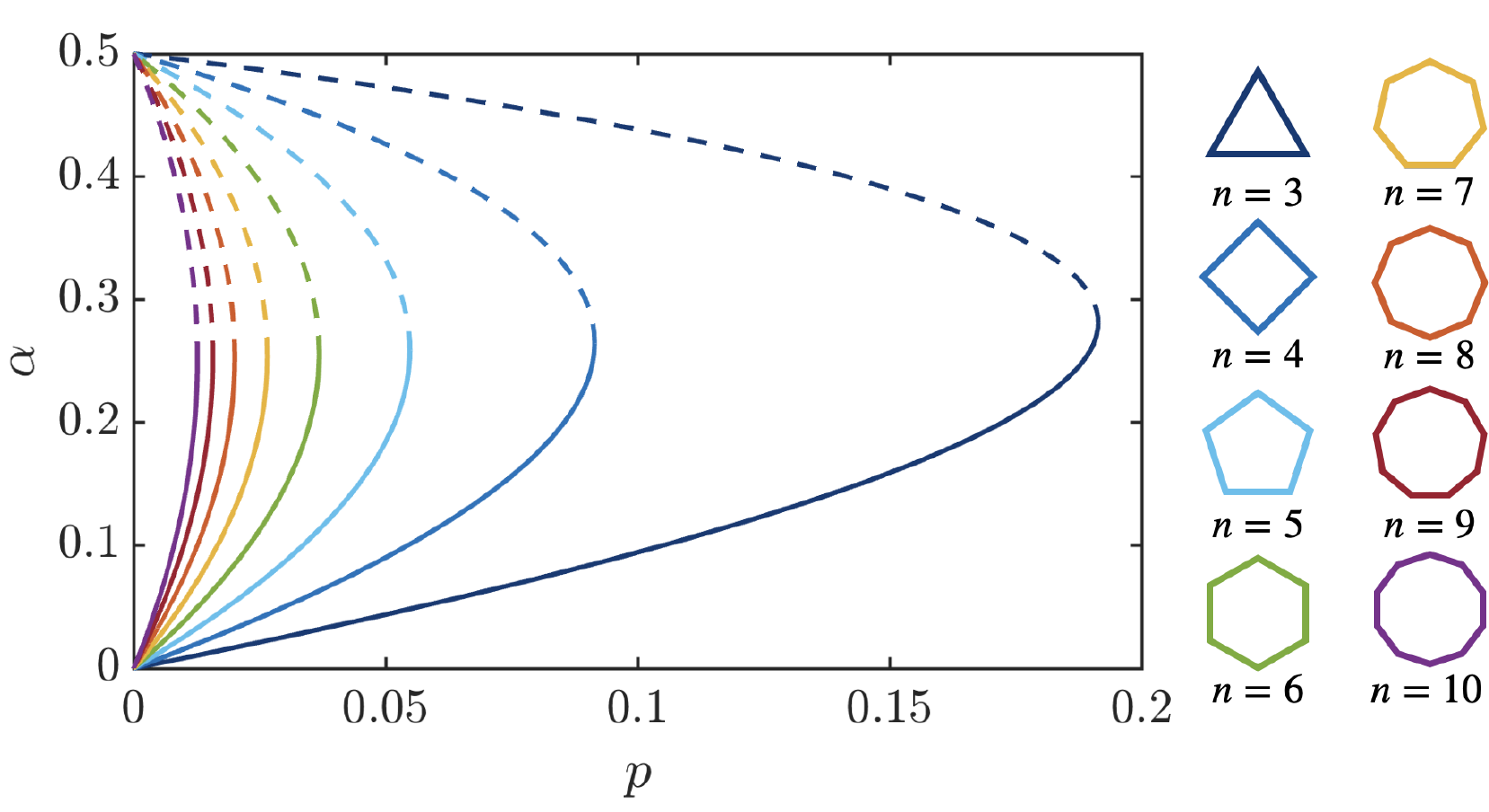}{Solutions of (\ref{eq:ode4_homothetic}): stable (solid) and unstable (dashed) branches of $\alpha=R/a$ in the function of $p$. Each stable branch point is an $n,\alpha,p$ parameter combination that results in a homothetic evolution. The geometries corresponding to $R/a=0$ for different $n$ values appear on the right.}

\section{Conclusions and discussion}
In the present work, we aimed to construct a mathematical model that facilitates our understanding of the morphology of fragment shapes from the earliest phase of abrasion. We derived a geometrically motivated ordinary differential equation (ODE) called the collisional polygon model to investigate the evolution of n-fold symmetric polygonal fragment shapes in two dimensions. The construction of the equation was inspired by the more compound collisional abrasion model proposed by Bloore. Although our model is a radically simpler construction, implying the idea of the collisional frequency dependent abrasion speed, it successfully adopts the key concepts of Bloore's partial differential equation (PDE).

The presented analytical approach reveals the underlying kinematics of the collisional abrasion of polygonal fragments with circular abraders of different sizes. The evolution provided by our model is compared to the results of Bloore's PDE, and a convincing qualitative and quantitative agreement was found in the case of fragments with $n=3,4$ and $5$.

Then we investigated the equation under different geometrical constraints to simulate various natural environments and specific evolution. Besides the well-known basic cases such as sandblasting, we deduced the conditions of stationary radii and homothetic evolution during the abrasion and discussed the effects of multiple abrader sizes. Providing a deeper, theoretically well-supported insight into the abrasion process, these findings reveal the driving factors of the evolution of fragment shapes.

Finally, the model was further extended to non-circular abraders, then the conditions and the interesting consequences of the self-dual case were presented. We showed that a non-monotonic evolution of the radius can be achieved in the self-dual and mixed cases. It was also shown that for some $n, R, a$ parameter combinations, it is possible to add dust to the system such that the shape evolves homothetically. Moreover, the homothetic solution is stable and attractive. 

Our findings of a non-monotonic evolution qualitatively outline a possible explanation of why natural fragments preserve their initial geometric features for a very long time, even after losing the majority of their initial masses.

\section{Acknowledgement}
Project no. TKP-6-6/PALY-2021 has been implemented with the support provided by the Ministry of Culture and Innovation of Hungary from the National Research, Development and Innovation Fund, financed under the TKP2021-NVA funding scheme. This research was supported by the NKFIH Hungarian Research Fund Grant 134199 is kindly acknowledged.


\bibliography{apssamp}

\end{document}